\documentclass[floats,aps,twocolumn,showpacs,amsmath,amssymb,superscriptaddress]{revtex4}
\usepackage{graphicx}
\usepackage{color}
\usepackage{amsmath}
\usepackage{dcolumn}
\usepackage{bm}

\input epsf
\def\k{{\bf k}}

\def\q{{\bf q}}
\begin{document}
\newcommand{\ltwid}{\mathrel{\raise.3ex\hbox{$<$\kern-.75em\lower1ex\hbox{$\sim$}}}}
\newcommand{\gtwid}{\mathrel{\raise.3ex\hbox{$>$\kern-.75em\lower1ex\hbox{$\sim$}}}}
\newcommand{\BSCCO}{{Bi$_2$Sr$_2$CaCu$_2$O$_8$ }}

\title{Relating STM, ARPES, and Transport in the Cuprate Superconducting State}

\author{D.J.~Scalapino}
\email{djs@vulcan2.physics.ucsb.edu}
\affiliation{Department of Physics, University of
California\\ Santa Barbara, CA 93106-9530 USA}

\author{T.S.~Nunner}
\email{nunner@phys.ufl.edu}
\author{P.J.~Hirschfeld}
\email{pjh@phys.ufl.edu}
\affiliation{Department of Physics, University of Florida\\
Gainesville, FL 32611 USA}

\date{\today}

\begin{abstract}

We discuss a wealth of data from various types of experiments
which together suggest that the superconducting state of optimally
to overdoped  BSCCO-2212 can be well-described by the BCS theory with a
$d$-wave gap together with small-angle scattering from out-of-plane defects.
These include scanning tunnelling
Fourier transform spectroscopy observation of nanoscale
inhomogeneity in the local gap edge position, the narrowing of the antinodal
ARPES spectrum when BSCCO becomes superconducting, as well as the behavior
of the  microwave and thermal conductivities. We suggest that the large
amount of small-angle scattering in BSCCO can account for the differences
between the superconducting properties of BSCCO and YBCO.

\end{abstract}

 \maketitle



\section{Introduction}

Much of our understanding about the  one-electron properties of the
cuprate superconducting state derives from experiments on one
material, BSCCO-2212, which cleaves between two BiO layers to
reveal atomically flat surfaces suitable for scanning tunnelling
microscopy (STM) and angle-resolved photoemission spectroscopy
(ARPES).  Many properties of this system differ quite
substantially from the other most-studied cuprate, YBCO-123, and
it is important to understand the differences which can be
attributed to details of electronic structure and disorder
in order to extract universal  aspects of cuprate superconductivity.
In recent years, high-resolution STM experiments have opened up
new windows on local aspects of electronic structure, and raised
questions about the role of disorder \cite{Yag99,HM03,How01}.
In particular, they have
revealed inhomogeneities in electronic structure at biases around
the gap scale on length scales of 25-30\AA.  Whether these patchy
structures, whose distribution changes with doping, are a result
of phase separation in  a correlated electronic system \cite{EK94},
or merely
a reflection of local doping disorder, is still an open question.


  Measurements of the ARPES spectrum of \BSCCO also raised important
questions regarding the role of disorder. Experiments showed the
clear emergence of a $d_{x^2-y^2}$-like gap in the superconducting
state of BSCCO. In the region of the antinodeal fermi surface
these studies revealed a broadened normal state spectrum which was
found to sharpen dramatically in the superconducting state
\cite{She93,Har97,Cam99,DHS03}.
Abrahams and Varma \cite{AV00} suggested that a major component of the normal
state broadening arose from small-angle elastic scattering due to
out-of-plane disorder, while a smaller contribution arose from
dynamic inelastic scattering. While the inelastic scattering rate
was expected to decrease below Tc as the superconducting gap
opened, the origin of the  apparent collapse of the {\it elastic}
scattering rate was unclear.

There has been also a longstanding puzzle regarding the very
different microwave conductivity temperature dependence observed
in the YBCO-123 \cite{Hos99} and BSCCO-2212 \cite{Lee96}
 systems.   The much smaller size of
the maximum conductivity and the insensitivity to changes in GHz
microwave frequency in BSCCO suggest that it is a much dirtier
system than YBCO; on the other hand, the data exhibits a large
residual $T\rightarrow 0$ component and a peak  at very low
temperatures, in contradiction to the usual picture of the
conductivity of a dirty $d$-wave system. The thermal conductivity
\cite{And00},
on the other hand, appears more standard, with a peak in $\kappa(T)$
near $T_c$;  the relative contribution of phonons and electrons to
the thermal current is not directly known, however, so one can
make no empirically based statements about $\kappa_{el}$ directly
without a theory of the material.

In this paper, we present a short review of work on the
aforementioned spectroscopies of BSCCO-2212 which, when taken
together, suggest to us that accounting for the way BSCCO is doped
can allow one to construct a unified picture of this fascinating
material and resolve many of the remaining puzzles. The presence
of interstitial O dopants in the BiO layer, together with many
other types of defects away from the CuO$_2$ plane, make this
indeed a dirty material, we believe. On the other hand, the
distance of the defects from the CuO$_2$ plane allows for some
screening, resulting in a rather smooth and weak random potential
landscape experienced by quasiparticles moving in the plane  and
lead to small-angle scattering. In addition, in-plane disorder can
lead to strong (near-unitary) elastic scattering and dynamic
spin-fluctuations lead to inelastic scattering which is suppressed
below $T_c$. We now explore the consequences of this picture.

\section{STM Data}
\subsection{FT-STS Patterns}

\begin{figure}[h]
\includegraphics[width=1.\columnwidth,clip,angle=0]{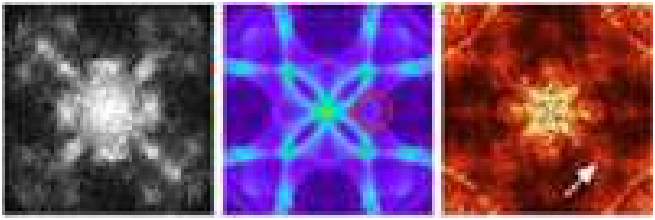}
\caption{a) Fourier transform STM image from \cite{HM03} on a
640\AA\ square of BSCCO-2212 surface at a voltage bias of
$-$14meV;. b) Theoretical $\rho(\q,\omega)$ at $-$14 meV with a
single point-like impurity. Red circle denotes position of
expected $\q_1$ peak from octet model. c) Many impurity simulation
from \cite{ZAH04} at $-$14meV with  0.2\% strong impurities
($V=30t$, with $t$ the near-neighbor hopping) and 8\% weak
extended impurities ($V=2t$, range $\sim a$). Arrow indicates
position of background feature.} \label{fig:STM1}
\end{figure}
 Among the many
fascinating aspects of recent Fourier transform - scanning
tunnelling spectroscopy  (FT-STS) experiments \cite{HM03,How01}
on BSCCO, we focus
here on the apparently mundane question of the size of the spots
observed in the Fourier pattern, generally taken to represent the
local density of states $\rho(\q,\omega)$ in the CuO$_2$ plane,
seen e.g. in the first panel of Fig.~\ref{fig:STM1}.  The various
spots in the picture have been claimed to correspond to the
 so-called ``octet vectors", the set of $\q$
connecting points of high density of states at the tips of the
curved elliptical contours of constant quasiparticle energy
\cite{HM03}. The single-impurity calculation of
$\rho(\q,\omega)$\cite{WangLee,Ting1imp}, although it reproduces
correctly the positions of many of  the octet vectors and their
dispersions, fails to correctly reproduce the weights of these
features, and misses some entirely, such as the $\q_1$ peak which
should fall inside the red circle in Fig.~\ref{fig:STM1}b. In
addition, the peaks are extremely sharp relative to experiment.
One might anticipate that adding more impurities would broaden the
peaks into the observed broad spots in Fig.~\ref{fig:STM1}a, but
the simplest calculation of many weak impurities gives
\cite{CSS03}

\begin{equation}
\delta \rho(\q,\omega) \simeq -V(\q) \mbox{Im
}\Lambda_3(\q,\omega)/\pi, \label{cap}
\end{equation}
where $V(\q)$ is the spatial Fourier transform of the
many-impurity potential, and $\Lambda_3$ is a response function of
the clean system with poles at the positions $\q$ of the octet
vectors similar to Fig.~\ref{fig:STM1}b.  Since $V(\q)$ is a
random function of $\q$ for a large number of impurities, we see
that as disorder increases there is no broadening of the
$\rho(\q,\omega)$ peaks, but rather  an increasing level of the
noise floor until the peaks are swamped.  A similar result can be
obtained for strong point-like impurities\cite{ZAH04}.  Thus the
only possibility to explain the widths of the peaks in these
experiments is to assume that the relevant impurities are quite
extended.  This is not a new concept: for some time studies of
$T_c$ supression in the cuprates have suggested that most of the
impurities away from the CuO$_2$ planes must act as small-angle
scatterers with little effect on $T_c$.  We believe that STM is
now seeing the local consequences of the presence of these
scatterers.

\subsection{Nanoscale Inhomogeneity}

\begin{figure}[h]
\begin{minipage}{.4\columnwidth}
\hspace{-.2\columnwidth}
\includegraphics[width=1.15\columnwidth,clip,angle=0]{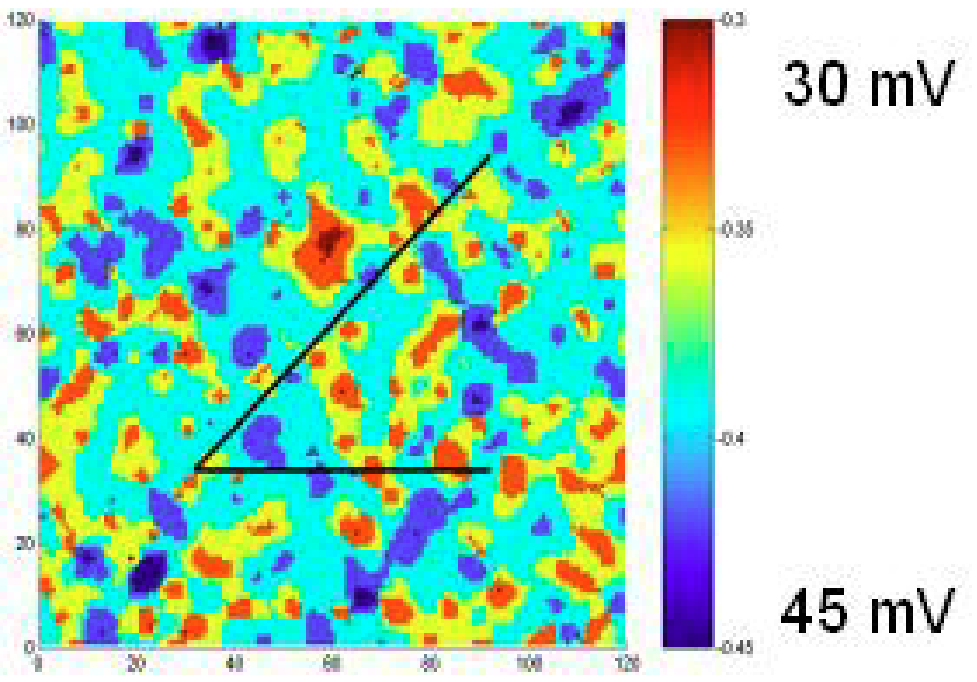}
\end{minipage}
\begin{minipage}{.49\columnwidth}
\hspace{-.1\columnwidth}
\includegraphics[width=1.1\columnwidth,clip,angle=0]{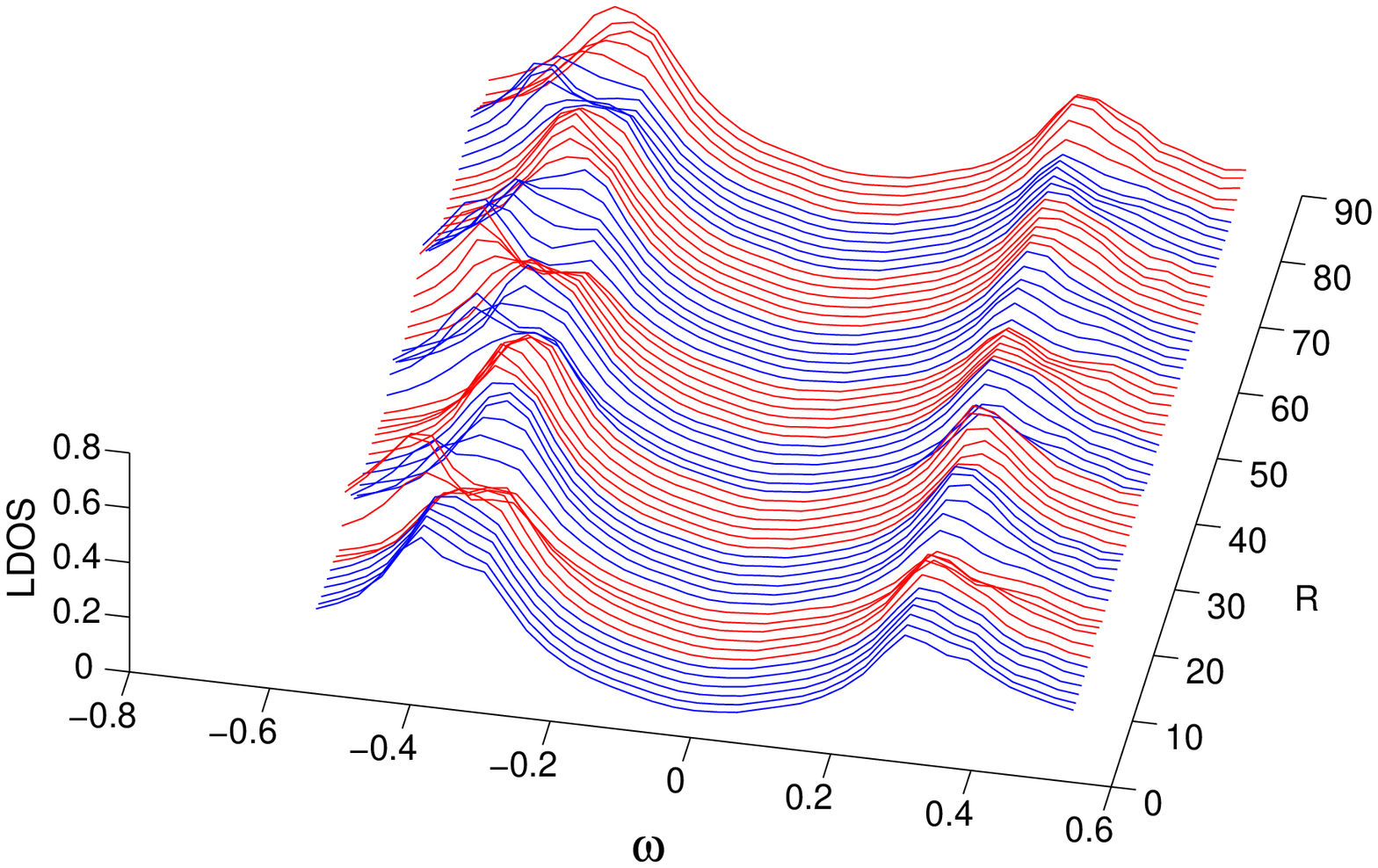}
\end{minipage}
\caption{Left: ``gap map" indicating position of negative bias
coherence peak from numerical solution of the Bogoliubov-de Gennes
equations in the presence of 3\% impurities with strength $V=.22t$.
Right: local density of states (LDOS) along a horizontal cut through
part of the map.}
\label{fig:STM2}
\end{figure}

The nanoscale inhomgogeneity at biases close to the bulk gap edge
observed in BSCCO STM experiments have been widely interpreted as
strong local fluctuations of the superconducting order parameter
with length scale of order 25-30 \AA, leading to many scenarios of
competing order in the cuprates.   While this may indeed be true,
and recent observations of charge ordering in ``patches"
characteristic of underdoped samples have lent support to this
point of view \cite{Hof02,Verxx}, it is worthwhile remembering
that the STM measures quasiparticle excitations rather than
superconducting order, and that quasiparticle interference effects
arising from the background disorder potential can also lead to
similar signatures. One of the most remarkable aspects of the STM
measurements is that they are completely {\it homogeneous} at
biases up to about 20 meV, suggesting that nodal quasiparticles
propagate freely without regard for the fluctuations which are
causing the nanoscale inhomogeneity at higher energies.  It is the
antinodal quasiparticles which appear disordered.  To illustrate
this point, in Fig.~\ref{fig:STM2} we show a ``gap map" depicting
the position of the `coherence peak" in a simulation (numerical
solution of the Bogoliubov-de Gennes equations \cite{ZAH04}) of
many impurities in a 2D $d$-wave superconductor. In this
particular simulation, only weak extended impurities were used. As
is well known, such defects do not cause resonant behavior at
subgap energies; hence the low-energy STM spectrum (also shown) is
completely homogeneous. On the other hand, the coherence peak is
seen to be quite distorted, with regions of both high and low gap.
In this particular case, the range of coherence peak position
oscillations is comparable to but smaller than observed in
experiment. However, it is noteworthy that this simulation was
performed assuming a {\it constant} order parameter; fluctuations
in the coherence peak position in the presence of disorder may
therefore involve interference effects which are not necessarily
directly related to the order parameter itself.  We expect this
may be particularly true in the optimally to overdoped materials,
where the concentration of the charge-ordered patches is
negligible.

\section{ARPES}

     The ARPES spectrum for the basic model described in the
introduction has been discussed in detail in \cite{ZHS04}. Here we review
this, focusing on the behavior of the small-angle elastic
scattering contribution. In the normal state just above $T_c$,
both the small-angle scattering and the spin-fluctuation inelastic
scattering lead to a broadening of the single particle spectral
weight. Now we know that the inelastic spin- fluctuation
scattering is suppressed as the temperature drops below $T_c$ and
the gap opens \cite{Hos99,DSH01}. The question is therefore, what happens to the
small-angle elastic scattering? In this case, the momentum
averaging over the Fermi surface, which leads to pairbreaking for
isotropic impurity scattering, is reduced. In particular, for
states near the Fermi surface, which are away from the nodes, one
approximately recovers Anderson's theorem \cite{And59} and the broadening due
to the forward elastic scattering is suppressed in the
superconducting state.

\begin{figure}[t]
\includegraphics[width=1.\columnwidth,clip,angle=0]{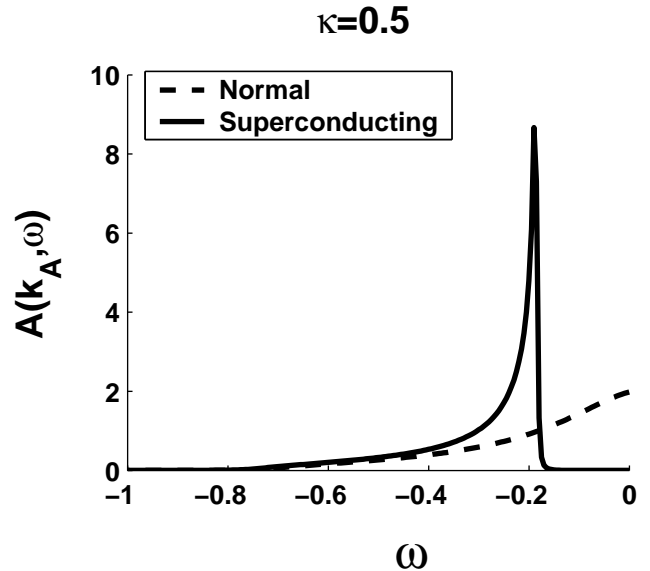}
\caption{Spectral function $A(\k,\omega)$ for forward elastic
scattering in the normal state (dashed) and superconducting state
(solid). Here, $\Gamma_0 (k_A) = \Delta (k_A) = 0.2t$ and the
scattering range parameter $\kappa= 0.5$. The frequency $\omega$
is measured in units of the near-neighbor hopping $t$. The inset shows the
upper quadrant of the Brillouin zone and the antinodal point ${\bf k}_A$.}
\label{fig:ARPESfig0}
\end{figure}

In \cite{ZHS04}, this physics was illustrated with a model
where impurity potentials decayed exponentially, $V(r)\propto
\exp(-\kappa r)$.   The limit $\kappa \rightarrow 0$ corresponds
to  pure forward scatterering, but realistic values are closer to
1 in units of $a^{-1}$.  Scattering was treated in the
self-consistent Born approximation.
        In the case where the elastic scattering is predominantly forward and
$\k$ is on the Fermi surface, the Green's function may be written as
[20,17]
\begin{equation}
        G(\k_F,\omega)={\omega\over\omega^2-\Delta(\k_F)^2+i \Gamma_0 (\k_F)
\sqrt{\omega^2-\Delta(\k_F)^2}}
\end{equation}
Here $\Gamma_0(\k_F)$ is the normal state elastic scattering rate
for $\k=\k_F$. Then for $\k_F$ at the antinodal point $\k_A$, shown in
the inset of Fig.~\ref{fig:ARPESfig0}, the
spectral weight varies as
\begin{equation}
        A(\k_A,\omega) ={\Delta(\k_A)\over \pi\Gamma_0(\k_A)
        \sqrt{\omega^2-\Delta(\k_A)^2}}
\end{equation}
while in the normal state

\begin{equation}
  A(\k_A,\omega)={\Gamma_0(\k_A)\over\omega^2+\Gamma_0(\k_A)^2}
\end{equation}
Figure \ref{fig:ARPESfig0} shows the normal state spectral weight (dashed)
and the superconducting spectral weight (solid) for $\omega < 0$ with $\Gamma_0
({\bf k}_A)=\Delta_0$. The ARPES
intensity would have a fermi factor multiplying $A(k, \omega)$ which would
round off the dashed curve leaving a broad response in the normal state.
One clearly sees the suppression of
the broadening in the superconducting state. Thus, a sharp spectral gap
feature in the ARPES spectrum for ${\bf k}_A$ at
temperatures below $T_c$ need not be in conflict with having a broad
spectrum in the normal state.


\section{Microwave Conductivity}

\begin{figure}[t]
\includegraphics[width=.9\columnwidth,clip,angle=0]{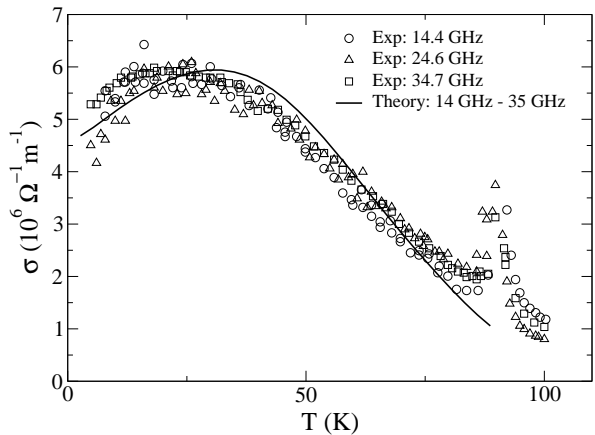}
\caption{Microwave conductivity of BSCCO-2212.  Data at 3
frequencies from \cite{Lee96}, theoretical curve from
\cite{NHun}.  Inelastic scattering rate calculated within
spin-fluctuation framework adopted from \cite{QSB94}, elastic
rate calculated using 11\% weak impurities with scattering
parameters $V_1=3t$, $V_2=0.4t$, $V_3=0.2t$ and 0.05\% unitarity
limit scatterers.  } \label{fig:muwave}
\end{figure}

Treating extended impurities in the calculation of transport is
complicated by the necessity of including vertex corrections to
the current-current correlation functions.  These are known to
vanish at $q=0$ for point-like scatterers in the $d$-wave state,
but are nonzero if the scattering is anisotropic.  This technical
difficulty was circumvented by Durst and Lee \cite{DL00}, who
linearized the quasiparticle dispersion at the node, and
parametrized the anisotropic elastic scattering potential
$V_{\k\k'}$ by three independent amplitudes $V_\alpha$, $\alpha=$
1, 2, 3 for a quasiparticle near one node of the $d$-wave order
parameter to scatter to all four nodes. This treatment renders the
problem finite-dimensional, and predicts the breakdown of the
universal microwave conductivity at $T\rightarrow 0,~
\omega\rightarrow 0$. In this case, the $T\to0, \omega\to0$ limit
of the conductivity is given by
\begin{equation}
\sigma_0 = {e^2\over \pi^2}{v_F\over v_g} \left( {V_1^2 + 2 V_2^2
+ V_3^2\over 2 V_2^2 + 2 V_3 ^2} \right),
\end{equation}
where $v_F$ and $v_g$ are the Fermi and gap velocities,
respectively.  In the isotropic limit, the universal value
${e^2\over \pi^2}{v_F\over v_g}$ (which turns out to be small in
the cuprates, of order $0.1-0.3$ $\sigma(T_c)$) is recovered.
Anisotropic scattering (unequal $V_\alpha$'s) implies a value of
$\sigma_0$  larger than the universal value, and the conductivity
diverges in the forward scattering limit $V_2=V_3=0$, $V_1\ne 0$.
In BSCCO, the most striking aspect of the measured microwave
conductivity, as shown in Fig.~\ref{fig:muwave}, is in fact the
apparent large finite value of $\sigma_0$.

 Nunner and Hirschfeld \cite{NHun} recently extended this technique to
finite frequencies and temperatures, treating strong pointlike
impurities within the $t$-matrix approximation, weak extended ones
within the Born approximation, and evaluating the total vertex
function. Although this technique is most suited for low $T$, and
expected to yield  only semiquantitative results near $T_c$,
  good agreement
is obtained  (Figure \ref{fig:muwave}) with a model consisting of
inelastic scattering from spin fluctuations identical to that used
to explain data on YBCO, plus roughly the same percentage of
strong and weak extended scatterers that was used to fit the STM
data.
 Calculations for YBCO show that good fits can
be obtained over the entire range of frequencies and temperatures
if the scatterers are assumed to be strong, but have
a small scattering range. Thus the microwave
data confirms the general picture that the defect type and
spatial distribution are quite different in YBCO and BSCCO crystals.

\section{Thermal Conductivity}

\begin{figure}[h]

\begin{minipage}{.53\columnwidth}
\includegraphics[width=.99\columnwidth,clip,angle=0]{ThermCondExpUnits.eps}
\end{minipage}
\begin{minipage}{.44\columnwidth}
\includegraphics[width=.99\columnwidth,clip,angle=0]{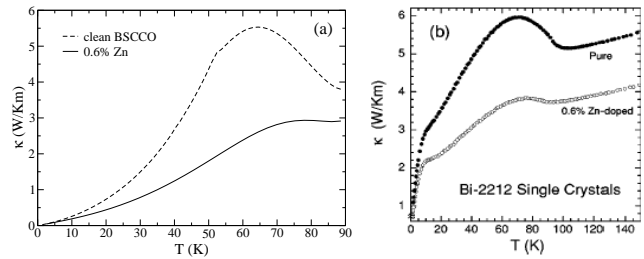}
\end{minipage}
\caption{Thermal conductivity of BSCCO-2212. (a)
Theoretical curve from \cite{NHun} for $\kappa_{el}(T)$.
Inelastic scattering rate calculated within spin-fluctuation
framework adopted from \cite{QSB94}, elastic rate calculated
using 11\% weak impurities with scattering parameters $V_1=3t$,
$V_2=0.01t$, $V_3=0.001t$ and 0.05\% unitarity limit scatterers
(upper curve). Lower curve has 0.6\% additional unitarity
scatterers. (b) Data on
optimally doped BSCCO-2212 single-crystal \cite{And00}.
Upper curve: nominally pure sample, lower curve: 0.6\% Zn.  } \label{fig:kappa}
\end{figure}

Finally we turn our attention to the thermal conductivity, where
the vertex corrections play a less important role, and small-angle
scattering processes contribute significantly to the transport
lifetime. While these corrections actually vanish at zero
temperature, they contribute at nonzero $T$, although they remain
small. To demonstrate consistency with the discussion of the
previous experiments, we compare in Fig.~\ref{fig:kappa}
calculated results with the data of Ando {\it et.~al}
\cite{And00}.  The peak heights and normal state values of both
the nominally pure and Zn-doped sample are comparable to those
seen in experiment, although the theoretical calculation is for
only the electronic part of the conductivity \cite{NHun}.  From
these fits it appears that, in contrast to YBCO, the phononic
contribution to the thermal conductivity in BSCCO must be quite
small relative to the electronic contribution near $T_c$, whereas
it dominates the transport at temperatures around $10K$. The
reduced relevance of the vertex corrections for this quantity
means that the large concentration of intermediate strength
scatterers broadens quasiparticle states and leads to a
temperature dependence similar to the standard ``dirty $d$-wave"
model.

\section{Conclusions}

Here we have argued that small-angle elastic scattering from
out-of-plane impurities can alter the electronic properties of the
cuprate superconductors and that this can account for differences
between BSCCO and YBCO. The weak Van der Waals coupling between
the BiO layers in BSCCO, which allow it to cleave so nicely
for ARPES and STM measurements, also provides a region in which
disorder can occur leading to small-angle elastic scattering of
the carriers in the CuO$_2$ layers. Here we have argued that when this
small-angle scattering is taken into account, along with the usual in-plane
impurity and inelastic spin-fluctuation scattering, one can understand the
differences between the superconducting properties of BSCCO and YBCO.

\acknowledgments

The authors thank W.A. Atkinson, L.-Y.~Zhu and T.~Dahm for
numerical calculations and  enlightening discussions. Partial
support was provided by ONR N00014-04-0060 (PJH), NSF-DMR02-11166
(DJS), and the A.~von Humboldt Foundation (TSN).

\end{document}